# Structure, Properties, and Disorder in the New Distorted-Hollandite PbIr$_4$Se$_8$

Benjamin A. Trump[a,b] and Tyrel M. McQueen[a,b,c]

[a]*Department of Chemistry, Johns Hopkins University, Baltimore, MD 21218, United States*
[b]*Department of Physics and Astronomy, Institute for Quantum Matter, Johns Hopkins University, Baltimore, MD 21218, United States*
[c]*Department of Material Science and Engineering, Johns Hopkins University, Baltimore, MD 21218, United States*

Abstract

The synthesis and physical properties of the new distorted-Hollandite PbIr$_4$Se$_8$ are reported. Powder X-ray diffraction and transmission electron microscopy show that the structure consists of edge- and corner-sharing IrSe$_6$ octahedra, with one-dimensional channels occupied by Pb. The structure contains Se-Se anion-anion bonding, leading to an electron count of Pb$^{2+}$(Ir$^{3+}$)$_4$(Se$_2$)$^{2-}$(Se$^{2-}$)$_6$, confirmed by bond-valence sums and diamagnetic behavior. Structural and heat capacity measurements demonstrate disorder on the Pb site, due to the combination of lone-pair effects and the large size of the one-dimensional channels. Comparisons are made to known Hollandite and pseudo-Hollandite structures, which demonstrates that the anion-anion bonding in PbIr$_4$Se$_8$ distorts its structure, to accommodate the Ir$^{3+}$ state. An electronic structure calculation indicates semiconductor character with a band gap of 0.76(11) eV.

1. Introduction

The Hollandite structure, prototypically α-MnO$_2$, is a well studied family of materials for battery,[1–3] thermoelectric,[4,5] and even magnetic applications.[6,7] The formula can be better written as $M_xT_4O_8$, where $M$ = early lanthanides, alkali metals, or alkali earth metals, $x$ = 0-1, and $T$ = Mn, Mo, Ru, or Ir.[8–11] A similar compound, the pseudo-Hollandites, have an even more expansive series, $M_xT'T_4Ch_8$ where $M$ = Tl, In, Cd, Sn, Pb, alkali metals, or alkaline earth metals, $x$ = 0-1, $T,T'$ = Ti, V, or Cr, and $Ch$ = S, Se, or Te.[12,13] The structure of both series consists of double chains of edge-sharing $T$-O octahedra which corner-share with other double octahedral chains to form a framework structure containing large one-dimensional (1-D), with a cation $M$, occupying the site in the channels.

Despite the wealth of cations and transition metals which take this structure, only KIr$_4$O$_8$ and Rb$_{0.68}$Ir$_4$O$_8$ contain a 5$d$ transition metal,[9,14] which have recently attracted significant interest due to strong spin-orbit coupling which could lead to nontrivial behavior.[15] These relativistic effects having comparable energy scales with crystal field stabilization or electron correlations is expected to lead to exotic quantum or magnetic behavior.[16–18] Iridium in particular, especially iridium chalcogenides, has been heavily studied for these reasons.

Though some non-oxide iridium chalcogenides have been studied, they are primarily derivatives of the binary compounds IrS$_2$, IrSe$_2$, and IrTe$_2$.[19–22] Some recent work has been done on stoichiometric ternary Ir-Sn-Se compounds,[23,24] though none have yet looked at other stoichiometric ternary Ir-Se-X compounds.

Here we report the synthesis, structure, and some physical properties of the new compound PbIr$_4$Se$_8$. This structure is analogous, but structurally distinct, to Hollandite, which contains similar 1-D channels. As far as the authors are aware, this is the first non-oxide 5$d$ Hollandite. Due to the large size of the 1-D channels, and Pb lone-pair effects, considerable disorder is seen on the Pb site, evidenced by power X-ray diffraction (PXRD), selected area electron diffraction (SAED), X-ray pair distribution analysis (PDF), and heat capacity. PbIr$_4$Se$_8$ is diamagnetic, indicative of low-spin 5$d^6$ Ir$^{3+}$, and has semiconducting character, evidenced by heat capacity and band structure calculations.

2. Experimental

2.1 Preparation

Polycrystalline PbIr$_4$Se$_8$ was grown by placing Ir (Alfa Aesar 99.95%), Pb (Alfa Aesar 99.999%), and Se (Alfa Aesar 99.999%) in the stoichiometric ratio of (PbSe)$_{1.1}$(IrSe$_2$)$_2$, in a fused silica tube, for a total of 300 mg. Tubes were backfilled with 1/3 atm of Ar to minimize vaporization of Se. The tube was heated quickly to 500 °C, followed by a 50 °C/h ramp to an annealing temperature of 950 °C, and held for four days before quenching in water. The resulting boule was pulverized, and the heat sequence and quench were repeated a second time. This resulted in shiny, silver PbSe micro-rings as well as loose, gray powder. After PbSe was mechanically removed, the resulting phase pure powder was used for all physical property and characterization methods. Attempts to target a myriad of alternate stoichiometries at several lower temperatures all led to the same PbIr$_4$Se$_8$ material with varied amounts of PbSe/IrSe$_2$ impurities. Specifically, attempts with less Pb and Se always included non-phase separating IrSe$_2$ impurities.

2.2 Characterization

Laboratory powder X-ray diffraction (PXRD) patterns were collected using a Cu K$_\alpha$ ($\lambda_{avg}$ = 1.5418 Å) on a Bruker D8 Focus diffractometer with a LynxEye detector. Peak searching and LeBail refinements were used for phase identification and initial lattice parameter estimates in TOPAS (Bruker AXS). Simulated annealing was then used to estimate atomic positions. Finally, Rietveld refinements determined precise atomic positions and lattice parameters, using TOPAS and GSAS-II[25] respectively. To verify the choice of lattice parameters and spacegroup, transmission electron microscopy (TEM) was used with a Phillips CM300 atomic resolution transmission electron microscope equipped with a field emission gun with an accelerating voltage of 300 kV. A CCD camera (bottom mounted Orius camera) was used to collect a tilt series of selected area electron diffraction (SAED) images, tilting from the [001] direction to the [$\bar{2}$ $\bar{1}$ 3] direction. Structures were visualized using VESTA.[26]

Synchrotron X-ray diffraction for pair distribution analysis (PDF) was collected at the beamline 11-ID-B at the Advanced Photon Source, Argonne National Laboratory with an X-ray wavelength of 0.2112 Å. A CeO$_2$ standard was used to estimate the resolution of the instrument. Data was reduced using Fit2D[27] and PDFgetX2[28] and finally analyzed using PDFgui[29].

Heat capacity and magnetization were collected on a cold-pressed pellets in a Physical Property Data Measurement system (PPMS, Quantum Design), for 1.8 K ≤ $T$ ≤ 300 K. Heat capacity was measured using the semiadiabatic pulse technique, with three repetitions at each temperature. Magnetization was measured with $\mu_0 H$ = 1 T.

2.3 Calculations

Electronic and band structure calculations were performed on $Ir_4Se_8$ using density functional theory (DFT) with the local density approximation (LDA) utilizing the ELK all-electron full-potential linearized augmented plane-wave plus local orbitals (FP-LAPW+LO) code.[30] Calculations were conducted both with and without spin-orbit coupling (SOC) using a 4 X 18 X 6 $k$-mesh, with the experimentally determined unit cell.

3. Results and Discussion

3.1 Structure of $PbIr_4Se_8$

Figure 1a shows the room-temperature laboratory PXRD data for $PbIr_4Se_8$ with a corresponding Rietveld refinement using spacegroup $C2/m$(12). Crystallographic parameters are shown in Table 1. Additional spacegroups were tested, and though Hamilton $R$ ratio tests[31] preferred a lower symmetry $C2$(5) cell with 90% confidence, $\chi^2$ ratios[32] only had 60% confidence for the $C2$ cell, hence the higher symmetry, $C2/m$, was chosen. Additionally, tests using ADDSYM in PLATON[33] did not find any additional symmetry.

SAED patterns, oriented close to the [$\overline{12}$ $\overline{5}$ 39] direction and along the [0 0 1] direction, are shown in Figures 1b and 1c respectively for $PbIr_4Se_8$, while the inset displays the respective perpendicular planes. Figure 1b displays diffuse scattering, or streaking, in the [$\overline{11}$ 1 4] direction. This SAED pattern is an example of the collected tilt series, which also displays streaking in the [$\overline{3}$ 0 1] and [6 0 1] directions. The SAED pattern in Figure 1c contains (0 $k$ 0) reflections, however the perpendicular direction does not consist of ($h$ 0 0) reflections alone. This is because the plane perpendicular to the [0 0 1] direction is not equivalent to the (0 0 1) plane, thus it is expected that the [0 0 1] direction has (0 $k$ 0) reflections perpendicular to (5$h$ 0 2$h$) reflections. The lattice parameters from SAED are within 10% different of those reported in Table 1. The streaking in some SAED patterns, however, hints at disorder within the structure.

The proposed structure of $PbIr_4Se_8$ is shown in Figure 2a, which is comprised of double chains of edge-sharing $IrSe_6$ octahedra which corner-share with other $IrSe_6$ double octahedral chains to form a framework structure containing 1-D channels. Pb resides in the 1-D channels in a $PbSe_8$ dual gyrobifastigium (di-rhombic prisms). The 1-D channels in PbIr$_4$Se$_8$ appear distorted in comparison to Hollandite, with Se-Se anion-anion bonding and a Se-Se bond distance of 2.50(2) Å in the smaller 1-D channels. This sort of anion-anion bonding is well known to occur in Ir chalcogenides[23,34–37] and leads to a formal electron count of Pb$^{2+}$(Ir$^{3+}$)$_4$(Se$_2$)$^{2-}$(Se$^{2-}$)$_6$, which is in good agreement with bond valence sums for an average Pb position (1.94(2)), and the material exhibiting diamagnetic behavior (i.e. Ir is low-spin $d^6$).

Despite the model fitting all observed peaks, with no evidence of impurities, the fit is visibly imperfect, highlighted by the insets in Figure 1a. These insets demonstrate that while some peaks are sharp, others are appreciably broadened. This peak broadening is consistent with the streaking seen in the SAED patterns, and indicates some degree of disorder in the material, however even $P$1 LeBail refinements do not visibly, or statistically, improve the fit. Additionally occupancies refined within 2% of nominal values and refinements which included strain (isotropic and anisotropic), anisotropic size, and anisotropic thermal parameters did not statistically improve the fit. The structure in Figure 2a gives a clue for this disorder, as the Pb site is in the center of large 1-D channels. Given the large size of these channels, and that even the shortest Pb-Se distance is greater than 3 Å, it is not surprising that there would be disorder on these sites. Likewise, Pb is well known to be stereochemically active from lone-pair effects,[38,39] which leads to disorder. This disorder would give rise to not only streaking in certain SAED patterns, but also would appreciably broaden any reflections whose intensities result from mainly Pb, while reflections resulting mainly from Ir or Se would not be affected. This same type of peak broadening was also seen in the PXRD data for the Hollandite Rb$_{0.68}$Ir$_4$O$_8$, which has disorder on the Rb site.[14] The most reasonable displacements involves Pb displacing to six possible positions, in the $xz$ and $y$ directions, along six of the Pb-Se directions. These directions are consistent with the directions of the observed streaking in SAED patterns. The displaced Pb sites gives an average bond valence sum of 2.04(2), which is closer to the ideal value of 2.00. The resulting atomic positions are shown in Table 1.

Given the streaking in SAED patterns and the imperfect Rietveld refinement, X-ray pair distribution analysis (PDF) was conducted, shown in Figure 3. The refined model is very similar to that for the Rietveld refinement - the only notable position difference is that Pb1 $y$ = 0.210(5) instead of $y$ = 0.345(2), corresponding to a small change in the direction and magnitude of the Pb off-centering. The thermal parameters are also more reasonable with $U_{iso}$ = 0.00508(3) Å$^2$, 0.00257(3) Å$^2$, and 0.00943(4) Å$^2$ for Pb, Ir, and Se respectively. The resulting Pb displacements are shown in Figure 2b. Though the PDF refinement fits remarkably well at short distances, at larger $r$ the fit is not as perfect, similar to lone-pair active Bi$_2$Ti$_2$O$_7$[40]. This is indicative of short range order and long range disorder, which is expected from the diffuse scattering seen in the SAED patterns, and K$_{1-x}$Ir$_4$O$_8$ is likely similar[41]. Future neutron pair distribution function studies are necessary to resolve the precise nature of this short range order.

3.2 Structural Similarities to Other Hollandites

The canonical Hollandite structure, α-MnO$_2$, shown as TlMn$_4$O$_8$ in Figure 4a, contains two separate 1-D channels of different sizes. Large 1-D channels are occupied by cations (e.g. In, Tl, Pb, alkali metals, alkaline earth metals), while smaller 1-D channels (light blue in Figure 4a) are empty. The distorted-Hollandite, PbIr$_4$Se$_8$, also contains two channels (Figure 4b) with the large 1-D channels occupied by Pb. PbIr$_4$Se$_8$ is structurally distinct from α-MnO$_2$ however, as the smaller 1-D channel contains Se-Se anion-anion bonding, contracting the structure. In comparison, the pseudo-Hollandite again contains large 1-D channels occupied by cations, however the smaller 1-D channels are also occupied, but with transition metals instead (Figure 4c). In the case of PbIr$_4$Se$_8$ the anion-anion bonding in the smaller 1-D channels appears to be due to charge balancing, as non-oxide Ir-chalcogenides always appear to maintain Ir$^{3+}$, from IrS$_2$, IrSe$_2$, IrTe$_2$, to Ir$_2$SnSe$_5$, and more. [23,34–37] These all have analogous portions of MnO$_2$ polymorphs, coupled with anion-anion bonding. The pseudo-Hollandite is also structurally distinct due to charge balancing, as TlCr$_5$Se$_8$ contains Tl$^{1+}$ and Cr$^{3+}$. If the formula unit were that of the canonical Hollandite, TlCr$_4$Se$_8$, Cr would instead have to be a mixture of Cr$^{3+}$ and Cr$^{4+}$ in order to accommodate the Tl$^{1+}$ cation, just as the Mn oxidation state in α-MnO$_2$ alters upon addition of cations.

Though charge balancing alone may explain the difference in these three structures, they are also expected to be stable based on closest-packing arguments, as previously elaborated by Klepp and Boller.[12] Using closest-packing of cations and anions in the *ac* plane, as closest-packing of *T-Ch* octahedra in the *b*-direction, ideal lattice parameters can be determined. Though these lattice parameters vary as a function of different sized cations and anions, the ratios of lattice parameters remain roughly constant for a range of ion sizes. In Table 2 we make the same comparison with an expanded range of Hollandites and pseudo-Hollandites, including our distorted-Hollandite. For a direct comparison to the pseudo-Hollandite and distorted-Hollandite, the Hollandite phases are described in a monoclinic setting, instead of their tetragonal setting, using $a_{mono}$ = √2 $a_{tetra}$, $b_{mono}$ = $b_{tetra}$, and $c_{mono}$ = $a_{tetra}$ (shown as dashed line in Figure 4a). The pseudo-Hollandite unit cells are in excellent agreement with the ideal lattice parameter ratios, over a range of cations and anions.[12] The Hollandite unit cells have more deviation from the ideal lattice parameter ratios, but that may be due to enhanced cation mobility at room temperature.[12] Though the Hollandite unit cells deviate from the ideal lattice parameter ratios, they are in excellent agreement with each other, despite a large variety of cations and transition metals.

In comparison to the other Hollandites, the lattice parameter ratios for our distorted-Hollandite deviate even farther from the ideal Hollandite lattice parameter ratios. In particular the *a/b* ratio and *β* of PbIr$_4$Se$_8$ are much smaller, while the *c/a* ratio is still in good agreement with the ideal Hollandite lattice parameter ratios. Since the magnitude of the *b*-axis is defined by the closest-packing of octahedra, the significant difference is the *a*-axis, which is shorter than ideal. Analyzing the PbIr$_4$Se$_8$ structure in Figure 4b, the framework is compressed in the *a*-direction, consistent with the deviation from the ideal Hollandite case, very likely due to Se-Se anion-anion bonding. This must mean that it is energetically more favorable for iridium to maintain Ir$^{3+}$ and introduce anion-anion bonding, than to obey closest-packing. It is important to note that neither the KIr$_4$O$_8$ or Rb$_{0.68}$Ir$_4$O$_8$

Hollandites undergo this distortion, as it is less energetically favorable to form O-O anion-anion bonding due to the increased electronegativity of oxygen.

With this understanding, the role of the anion-anion bonding appears to be to allow iridium to be $Ir^{3+}$, it is just a question of whether it is energetically more favorable for the anions to share electrons, or for iridium to be in the 3+ state. In the case of sulfides, selenides, and tellurides, it appears that it is universally more favorable for Ir to be $Ir^{3+}$ with anion-anion bonding. The low-spin $5d^6$ state is exceptionally stable, thus doping such a state might incite exotic quantum behavior due to electron mobility on the anion framework, rather than changing the oxidation state of $Ir^{3+}$ to $Ir^{4+}$, such as in $Ir_{1-x}M_xTe_2$ ($M$ = Pd or Pt),[19–22] which host superconductivity.

These models also explain why the $M_xIr_4O_8$ Hollandites are metallic and paramagnetic,[9,14] while $PbIr_4Se_8$ is insulating and diamagnetic. The difference between these compounds is the oxidation state of iridium; in $MIr_4O_8$ iridium is in the $Ir^{3.75+}$ state, while in $PbIr_4Se_8$ iridium is in the $Ir^{3+}$ state. $Ir^{3+}$ is comprised of low-spin $5d^6$, with all of the $t_{2g}$ orbitals filled, while a low-spin $Ir^{3.75+}$ would have two filled $t_{2g}$ orbitals, and a degenerate $t_{2g}$ orbital with unpaired electrons, giving rise to paramagnetic and metallic behavior as observed in $MIr_4O_8$ compounds.

Most Rietveld refinements for Hollandite polymorphs (pseudo-Hollandites, Cryptomelane, Priderite, Psilomelane) which contains cations refines these cations to either significantly large isotropic thermal parameters, or highly anisotropic thermal parameters.[1–14,42–44] Some studies have analyzed these in great detail, from lone-pair displacements in $InCr_5S_8$ in purely the $b$ direction,[42] to theoretical investigations demonstrating it is more energetically favorable for Li to displace in the $ac$ direction in $LiMnO_2$.[45] Each compound appears to have its own degree of cation disorder, but the literature seems to show that both displacements due to small cations in large channels and lone-pair effects are possible; and both in the $b$ and $ac$ directions. This means that our model, which displaces Pb in both the $b$ and $ac$ directions, is likely realistic for our system, with the displacement being caused by a combination of lone-pair effects and the size of the channels. Though it would be ideal to characterize this new material with a metal in the channel that does not displace, it is no small challenge to find a metal with no lone-pair activity, is appropriately sized, and forms the same structure.

3.3 Heat Capacity

Resulting heat capacity data for $PbIr_4Se_8$ is shown in Figures 5a and 5b. Figure 5a shows a plot of $C_p/T$ vs $T^2$, with a linear fit to $C_p/T = \beta_3 T^2 + \gamma$, a low-temperature approximation, where $\beta_3$ represents the phonon contribution, and $\gamma$, the Sommerfeld coefficient, represents the electronic heat capacity. The value of $\gamma$ = 4.5(8) mJ mol$^{-1}$ K$^{-2}$ indicates a small, non-zero density of states of the Fermi level, due to some small number of states at the Fermi level. Although occupancies refined within 2% of nominal values, the structural distortion could obfuscate Se or Pb vacancies, which are common and would explain the non-zero Sommerfeld coefficient. Alternatively this could be due to the Fermi level lying on the edge of a valence band. Figure 5b shows the heat capacity for PbIr$_4$Se$_8$ as $C_p/T^3$ vs log $T$ to highlight acoustic and optic phonon modes.[46] Plotted this way Einstein (optic) modes appear to peak while Debye (acoustic) modes increase upon cooling until becoming constant. The data was fit by a model with one Einstein and two Debye modes, as well as an electronic contribution:

$$C_p/T^3 = E\,(\theta_E, T)/T^3 + D1\,(\theta_{D1}, T)/T^3 + D2\,(\theta_{D2}, T)/T^3 + \gamma/T^2$$

Where $\theta_E$ is the Einstein temperature and $\theta_D$ is the Debye temperature. The Einstein model approximates an optic mode by:

$$E(\theta_E, T) = 3sR \left(\frac{\theta_E}{T}\right)^2 \frac{\exp(\theta_E/T)}{[\exp(\theta_E/T) - 1]^2}$$

While the Debye model approximates an acoustic mode by:

$$D(\theta_D, T) = 9sR \left(\frac{T}{\theta_D}\right)^3 \int_0^{\theta_D/T} \frac{(\theta/T)^4 \exp(\theta/T)}{[\exp(\theta/T) - 1]^2} d\frac{\theta}{T}$$

Where $s$ is the oscillator strength and $R$ is the molar Boltzmann constant.[47] This model appears to be in excellent agreement with the data in Figure 5b, and each of the separate contributions are shown. An initial fit was attempted using the Debye temperature calculated from $\beta_3$, however this severely under-fit the data and a second Debye mode was added. The final fit parameters were $s_{D1}$ = 11.61(4), $\theta_{D1}$ = 363(2) K, $s_{D2}$ = 1.36(4), $\theta_{D2}$ = 100.6(9) K, $s_E$ = 0.259(4), and $\theta_E$ = 41.0(2) K, where the total number of oscillators adds up to 13.2(1), in good agreement with the total number of atoms per formula unit. The large Einstein mode is due to the Pb disorder, as similar effects are commonly seen in the literature.[38,39]

Table 3 shows a comparison of both Einstein mode and physical displacement magnitude for a variety of lone-pair active compounds. The Einstein energy among the pyrochlore Pb$_2$Ru$_2$O$_{6.5}$, the perovskite PbTiO$_3$, and the distorted-Hollandite PbIr$_4$Se$_8$ all have the same magnitude, though the value for PbIr$_4$Se$_8$ is roughly half. This could be due to the large 1-D channels, which would entropically decrease the Pb order, verifying that cation displacements in Hollandite polymorphs are not due to lone-pair effects alone. The magnitude of displacements across the perovskite PbTiO$_3$, distorted-Hollandite PbIr$_4$Se$_8$, and pseudo-Hollandite InCr$_5$Se$_8$ are in rough agreement as well, as all of these structure types have "pockets" large enough to allow this displacement, unlike the pyrochlore Pb$_2$Ru$_2$O$_{6.5}$, which has a much smaller displacement due to the significantly different structure.

3.4 Band Structure

The disorder of the Pb site precludes direct computation of the band structure. However, a reasonable approximation can be derived by computing the band structure for the host Ir$_4$Se$_8$ framework, and then applying the rigid band approximation to shift the Fermi level consistent with the inclusion of one $Pb^{2+}$ per formula unit. Such a band structure for $Ir_4Se_8$ is shown in Figure 6a, plotted using the Brillouin zone definition in Figure 6b. Introducing $Pb^{2+}$ into the $Ir_4Se_8$ framework shifts the Fermi level up, donating two electrons per formula unit. The dotted line in Figure 6a displays the new location of the Fermi level, derived from integrating the density of states. As $\gamma$ is proportional to the number of states at the Fermi level, the proposed Fermi level position is in good agreement with the small, non-zero $\gamma$, both due to the small number of states, and the curvature of the band. The 0.75(11) eV gap between the new Fermi level and the conduction band indicates semiconducting behavior which was observed experimentally as charging effects in the TEM beam.

4. Conclusion

The new $PbIr_4Se_8$ distorted-Hollandite is reported. It is a network of corner- and edge-sharing $IrSe_6$ octahedra, Se-Se anion-anion bonding in small 1-D channels, and Pb in large 1-D channels. Laboratory X-ray data and selected area electron diffraction demonstrate that the material is well described by spacegroup $C/2m$(12), however there is evidence for Pb disorder due to lone-pair effects and the large size of the 1-D channels. An Einstein mode in heat capacity measurements further confirms the Pb disorder, due to a small, non-zero Sommerfeld coefficient of $\gamma$ = 4.5(8) mJ mol$^{-1}$ K$^{-2}$. Band structure calculations further confirm the small non-zero $\gamma$, and demonstrate $PbIr_4Se_8$ is a semiconductor, with a band gap of 0.76(11) eV. The distorted-Hollandite $PbIr_4Se_8$ is structurally similar to other Hollandite polymorphs, but the 1-D channels are distorted to allow for Se-Se anion-anion bonding in the smaller 1-D channels, which accommodates Ir to be $Ir^{3+}$, despite deviation from close-packing. This is consistent with other non-oxide Ir chalcogenides which also exhibit anion-anion bonding, as well as diamagnetic and insulating behavior. Given the disorder of the Pb in the 1-D channels, $PbIr_4Se_8$ is a good candidate for thermoelectric materials. Similarly, as $PbIr_4Se_8$ is a Hollandite polymorph, it would also be a candidate for battery materials, especially given that the Se-Se anion-anion bonding may make the framework $Ir_4Se_8$ more stable.

Acknowledgements

This work was supported by NSF, Division of Materials Research, Solid State Chemistry, CAREER grant under Award DMR-1253562. BAT also acknowledges useful discussions with K. J.T. Livi. The research also benefited from beamline 11-ID-B of the Advanced Photon Source, a U.S. Department of Energy (DOE) Office of Science by Argonne National Laboratory under Contract No. DE-AC02-06CH11357. We also acknowledge K.E. Arpino for technical assistance.

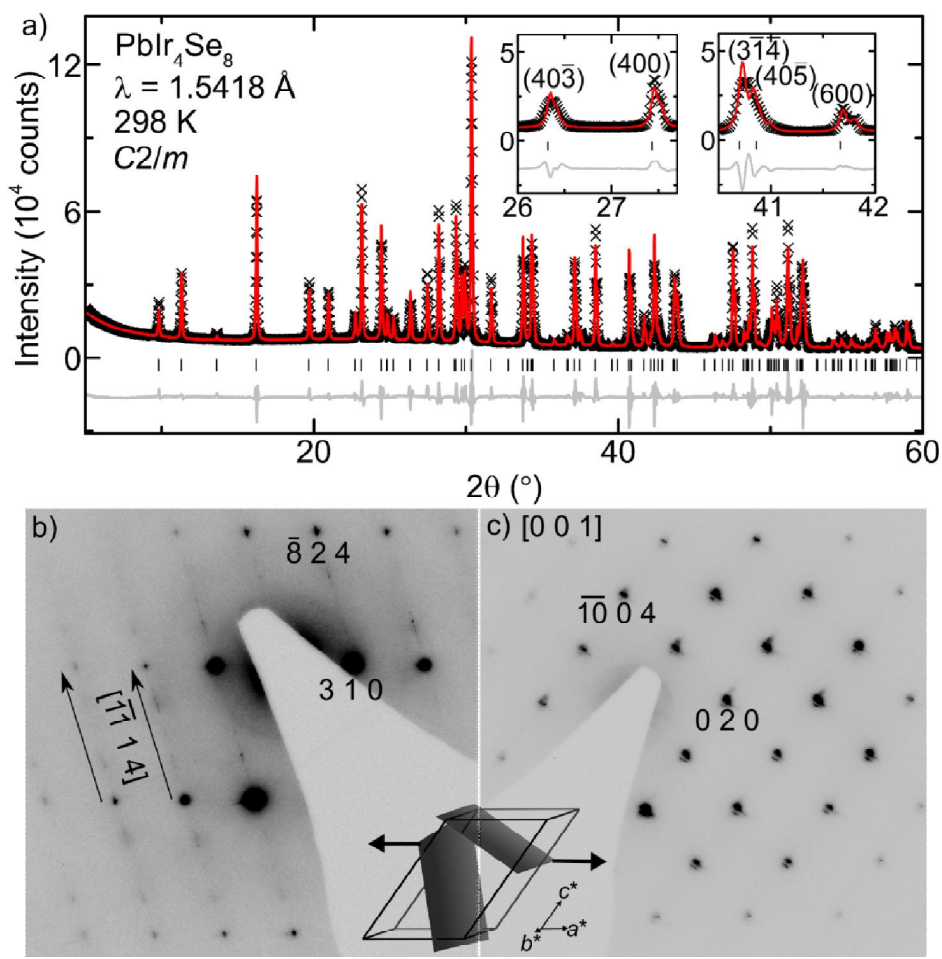

**Figure 1**. a) Laboratory powder X-ray diffraction for PbIr$_4$Se$_8$ shown as black X's, fit in red, and difference in gray. Insets demonstrate some peaks are appreciably broadened. b) Transmission electron diffraction close to the [$\overline{12}$ $\overline{5}$ 39] direction and c) along the [001] direction for PbIr$_4$Se$_8$. The inset displays the corresponding planes. Diffuse scattering (streaking) is seen in the [$\overline{11}$ 1 4] direction.

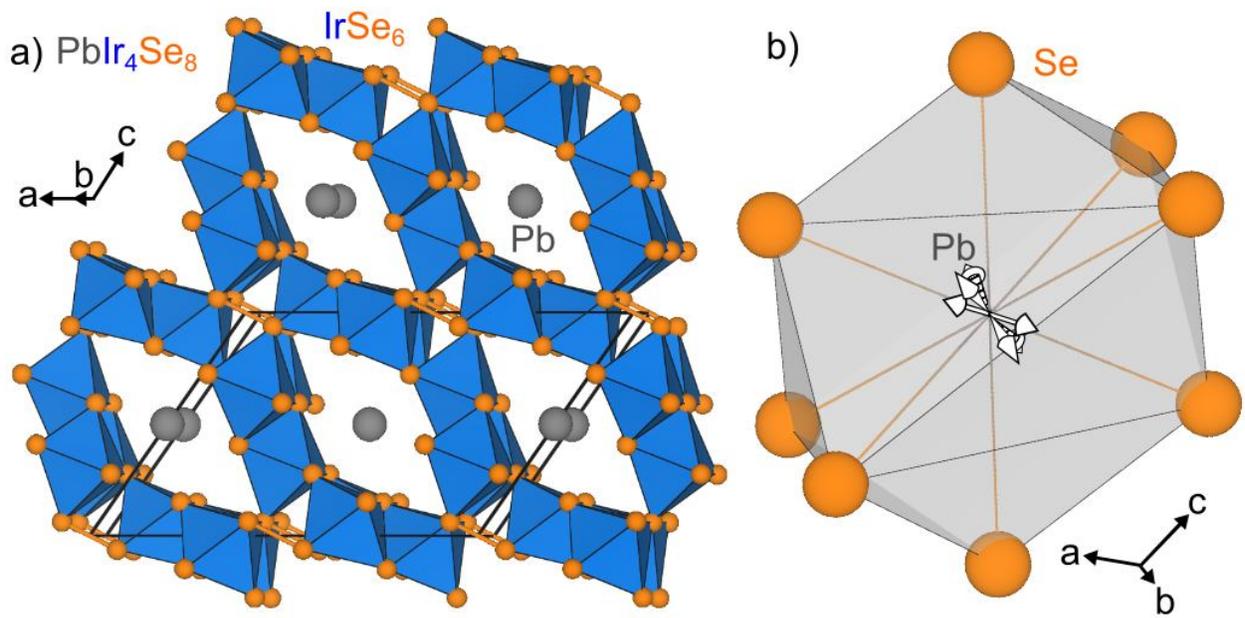

**Figure 2**. a) The PbIr$_4$Se$_8$ distorted-Hollandite structure. Pb positions shown are an average of sites. b) The PbSe$_8$ dual gyrobifastigium highlighting the direction and magnitude of the modeled Pb displacement. Pb shown in grey, Se in orange, and IrSe$_6$ octahedra in blue.

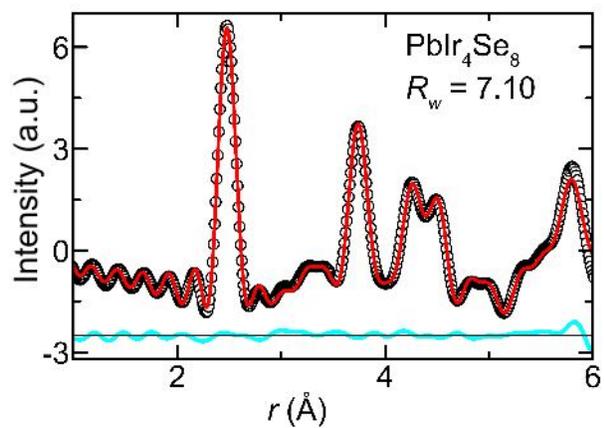

**Figure 3**. X-ray pair distribution analysis on PbIr$_4$Se$_8$ using the displaced Pb model.

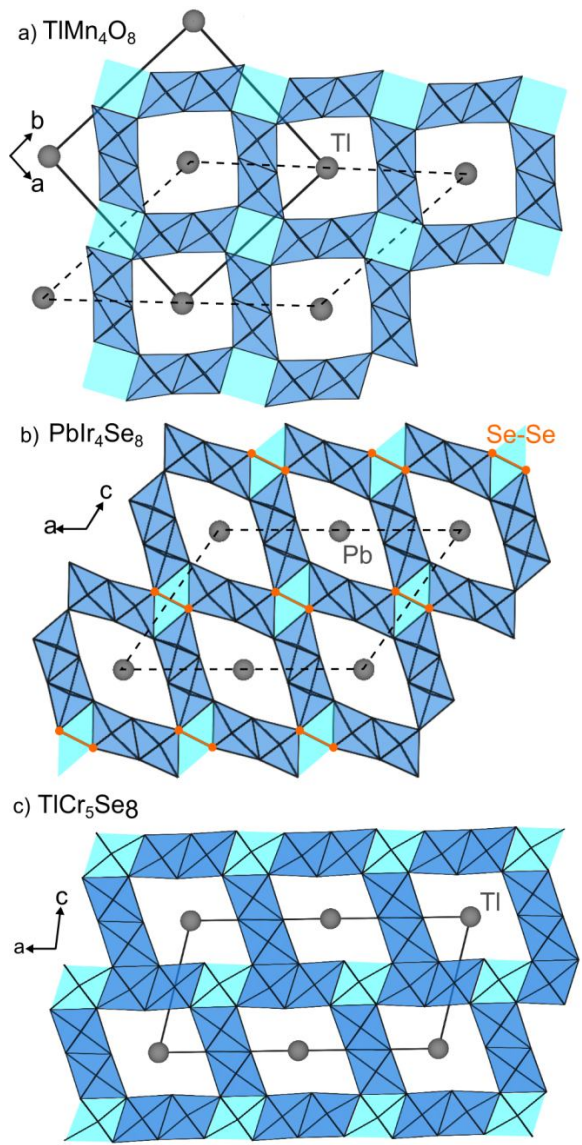

**Figure 4**. a) The MnO$_2$ hollandite structure (TlMn$_4$O$_8$), which contains both large 1-D channels occupied by cations (Tl) and smaller, empty 1-D channels. b) The PbIr$_4$Se$_8$ structure with Pb in large, distorted 1-D channels, and Se-Se anion-anion bonding in small 1-D channels. c) The TlCr$_5$Se$_8$ pseudo-Hollandite structure. Here the large 1-D channels are occupied by Tl while the small channels are occupied by Cr. Blue shading represents $TCh_6$ octahedra, light blue shading highlights the small 1-D channels, and dashed lines indicate alternative, comparative unit cells.

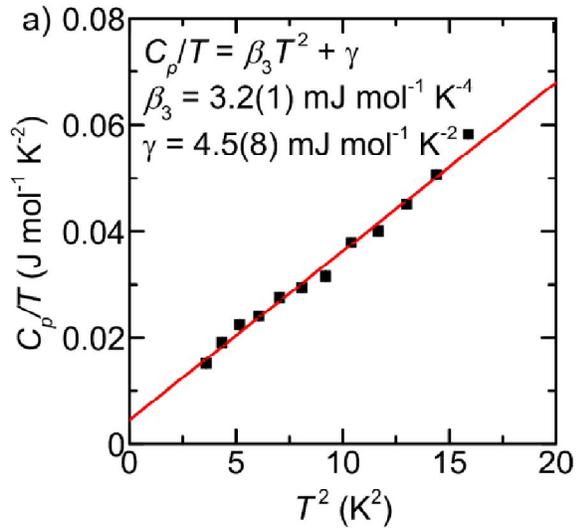

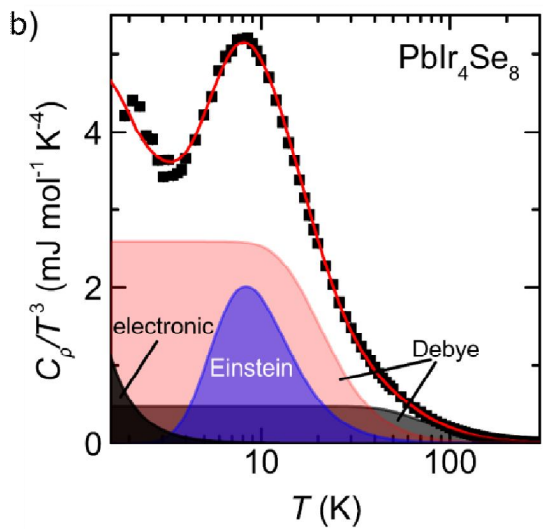

**Figure 5**. a) Heat capacity over temperature versus temperature squared. A non-zero Sommerfeld coefficient (γ) is seen. b) Heat capacity over temperature cubed versus temperature demonstrates an Einstein-like mode is clearly seen. Data is shown as black squares, with red lines as fits for both.

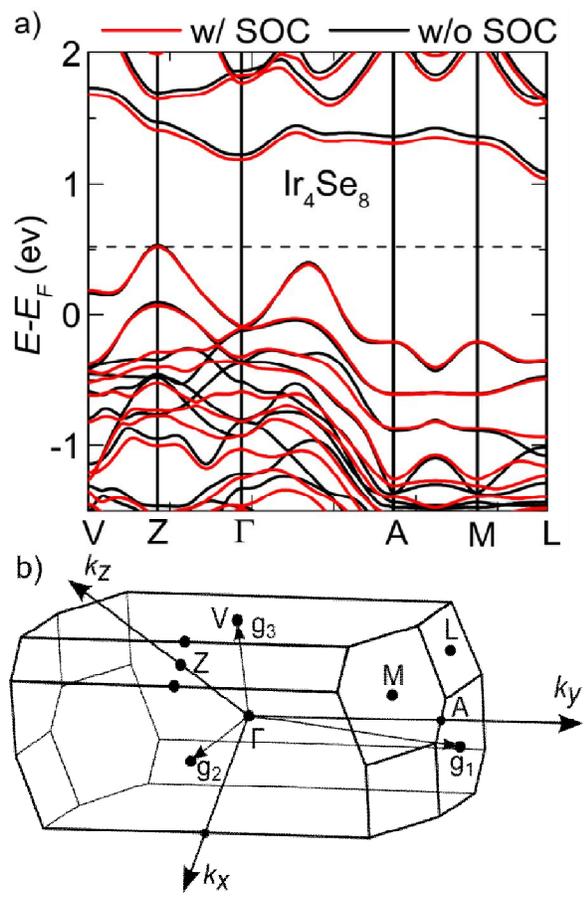

**Figure 6**. a) Band structure for the $Ir_4Se_8$ cages, without Pb, both with and without spin-orbit coupling (SOC). Dotted line shows the location of the Fermi level when Pb donates two electrons per formula unit. b) Brillouin zone for $PbIr_4Se_8$.

**Table 1.** Crystallographic fit parameters for PbIr$_4$Se$_8$ using spacegroup $C2/m$(12). All atoms are on the 4$i$:($x$ 0 $z$) site, except for Pb1 which is on the 8$j$:($x$ $y$ $z$) site. Occupancies on the two Pb Wyckoff sites were fixed to a total of 2 Pb per cell, all other occupancies were fixed at unity. Errors represent statistical uncertainties.

|  |  |  |  |  |  |  |  |  |
|---|---|---|---|---|---|---|---|---|
|  | $a$ (Å) | 15.901(3) |  | $V$ (Å$^3$) | 534.90(7) |  | $R_{wp}$ (%) | 8.23 |
|  | $b$ (Å) | 3.7300(1) |  | $\lambda$ (Å) | 1.5418 |  | $R_p$ (%) | 5.41 |
|  | $c$ (Å) | 11.035(4) |  | $T$ (K) | 296(1) |  | $R_F^2$ (%) | 5.87 |
|  | $\beta$ (°) | 125.190(8) |  |  |  |  | $\chi^2$ (%) | 1.97 |
| Ir1 | $U_{iso}$ (Å$^2$) | 0.0173(4) | Se2 | $U_{iso}$ (Å$^2$) | = $U_{iso}$(Se1) | Pb1 | $U_{iso}$ (Å$^2$) | 0.0214(5) |
|  | $x$ | 0.3670(2) |  | $x$ | 0.2200(5) |  | occ | 0.16667 |
|  | $z$ | 0.0422(3) |  | $z$ | 0.5924(6) |  | $x$ | 0.498(3) |
| Ir2 | $U_{iso}$ (Å$^2$) | = $U_{iso}$(Ir1) | Se3 | $U_{iso}$ (Å$^2$) | = $U_{iso}$(Se1) |  | $y$ | 0.345(2) |
|  | $x$ | 0.8454(2) |  | $x$ | 0.0956(4) |  | $z$ | 0.462(2) |
|  | $z$ | 0.6749(3) |  | $z$ | 0.0655(7) | Pb2 | $U_{iso}$ (Å$^2$) | = $U_{iso}$(Pb1) |
| Se1 | $U_{iso}$ (Å$^2$) | 0.0150(4) | Se4 | $U_{iso}$ (Å$^2$) | = $U_{iso}$(Se1) |  | occ | 0.16667 |
|  | $x$ | 0.1756(5) |  | $x$ | 0.4638(5) |  | $x$ | 0.030(1) |
|  | $z$ | 0.8503(6) |  | $z$ | 0.7227(7) |  | $z$ | 0.509(6) |

**Table 2.** Unit cell parameters and ratios of Hollandites, pseudo-Hollandites, and the new distorted-Hollandite for comparison. Ideal ratios are derived for a closest-packing model.

|  | $a$(Å) | $b$(Å) | $c$ (Å) | $\beta$ (°) | $a/b$ | $c/a$ |
|---|---|---|---|---|---|---|
| *pseudo-Hollandite* |  |  |  |  |  |  |
| TlV$_5$S$_8$[12] | 17.465 | 3.301 | 8.519 | 103.94 | 5.29 | 0.488 |
| TlTi$_5$Se$_8$[12] | 18.773 | 3.583 | 9.1065 | 104.13 | 5.24 | 0.485 |
| Rb$_{0.62}$Cr$_5$Te$_8$[43] | 20.367 | 3.902 | 9.605 | 104.39 | 5.22 | 0.472 |
| **Ideal**[12] |  |  |  | **103.26** | **5.10** | **0.484** |
| *Hollandite*[a] |  |  |  |  |  |  |
| Mn$_4$O$_8$[8] | 13.842 | 2.865 | 9.788 | 133.82 | 4.83 | 0.707 |
| Ba$_{0.7}$Sn$_{2.6}$Cr$_{1.4}$O$_8$[12] | 14.728 | 3.108 | 10.012 | 134.37 | 4.47 | 0.680 |
| Nd$_{2/3}$Mo$_4$O$_8$[10] | 13.999 | 2.940 | 9.899 | 134.68 | 4.76 | 0.707 |
| KRu$_4$O$_8$[11] | 13.953 | 3.131 | 9.866 | 133.63 | 4.46 | 0.707 |
| Rb$_{0.68}$Ir$_4$O$_8$[14] | 14.284 | 3.149 | 10.100 | 133.67 | 4.54 | 0.707 |
| **Ideal**[12] |  |  |  | **133.31** | **5.20** | **0.680** |
| *distorted-Hollandite* |  |  |  |  |  |  |
| PbIr$_4$Se$_8$[this work] | 15.901 | 3.730 | 11.035 | 125.19 | 4.26 | 0.694 |

[a]For direct comparison Hollandite phases are reported in a monoclinic setting (instead of tetragonal).

**Table 3**. Comparison of Einstein modes and displacements ($\delta$) for compounds with lone-pair active cations.

| | s | E (meV) | $\delta$ (Å) |
|---|---|---|---|
| $PbTiO_3$[36] | 0.4(1) | 5.5(7) | 0.474[43] |
| | 1.5(5) | 8(1) | |
| $Pb_2Ru_2O_{6.5}$[35] | | 6(2) | 0.020(4) |
| $PbIr_4Se_8$[a] | 0.259(4) | 3.53(2) | 0.55(15) |
| $InCr_5S_8$[37] | | | 0.363 |

[a]This work.